\documentclass[twocolumn]{aastex631}

\usepackage{amsmath}

\graphicspath{{./}{figures/}}

\newcommand{\OII}{[\ion{O}{2}]}
\newcommand{\OIII}{[\ion{O}{3}]}
\newcommand{\NeV}{[\ion{Ne}{5}]}

\begin{document}

\title{Tracing Star Formation in Quasar Hosts via \OII\ $\lambda$3727: A Kinematically Consistent Approach}

\author{Liang Wu$^*$}
\affiliation{Department of Optoelectronic Engineering, College of Mechanical and Electrical Engineering, Chizhou University, Chizhou, Anhui, 247000, China}
\affiliation{Department of Astronomy, School of Astronomy and Space Science, University of Science and Technology of China, Hefei, Anhui, 230026, China}

\author{Jun-Xian Wang$^*$}
\affiliation{Department of Astronomy, School of Astronomy and Space Science, University of Science and Technology of China, Hefei, Anhui, 230026, China}
\affiliation{
College of Physics, Guizhou University, Guiyang, Guizhou 550025, People’s Republic of China}

\author{Luis C. Ho}
\affiliation{Kavli Institute for Astronomy and Astrophysics, Peking University, Beijing 100871, China}
\affiliation{Department of Astronomy, School of Physics, Peking University, Beijing 100871, China}

\author{Junfeng Wang}
\affiliation{Department of Astronomy, Xiamen University, Xiamen, 361005}

\author{Zhicheng He}
\affiliation{Department of Astronomy, School of Astronomy and Space Science, University of Science and Technology of China, Hefei, Anhui, 230026, China}

\correspondingauthor{Liang Wu \& Jun-Xian Wang} 
\email{wul@mail.ustc.edu.cn, jxw@ustc.edu.cn}

\begin{abstract}
Measuring star formation in quasar host galaxies is crucial for understanding the coevolution of supermassive black holes (SMBHs) and galaxies, yet remains observationally challenging due to severe contamination from active galactic nucleus (AGN) emission. In this work, we present a new method to robustly isolate the AGN contribution to the [O II] $\lambda$3727 emission line in quasars, based on a kinematically consistent decomposition of [O II] and the high-ionization [Ne V] $\lambda$3426 line. We find that the [O II] emission in quasars is primarily dominated by star formation, with only a weak AGN contribution, and thus can be reliably used as a tracer of star formation in quasar hosts. Applying this technique to a large sample of Sloan Digital Sky Survey quasars, we derive mean SFRs as a function of bolometric luminosity. We find a tight correlation between mean SFR and luminosity. Further analysis, { assuming a constant dust extinction correction to \OII\ emission,} shows that luminosity is the primary parameter most strongly associated with star formation, rather than SMBH mass or Eddington ratio. This supports the scheme in which star formation and black hole accretion are closely linked through their common dependence on the cold gas supply.
\end{abstract}


\section{Introduction}


Measuring the star formation activity in AGN host galaxies provides a fundamental observational avenue for exploring the coevolution of supermassive black holes and their host galaxies.
Despite decades of research, the relationship between AGN activity and host galaxy star formation remains unclear, with observations yielding conflicting results (see \citealt{Harrison2017,Alexander2025} for reviews). 
These discrepancies likely arise from a combination of intrinsic physical diversity and observational uncertainties.

A major source of these uncertainties is the difficulty of reliably measuring star formation rates (SFRs) in large sample of AGN host galaxies.
Traditional tracers of recent star formation, such as H$\alpha$ and ultraviolet (UV), are generally unreliable in type 1 AGN.
In this context, the [O II] $\lambda$3727 line has emerged as a promising alternative, as it is strongly excited by young stars but only weakly produced in AGN narrow-line regions \citep[e.g.][]{kewley2004}. Nevertheless, its application as a star formation tracer in AGN hosts faces important limitations. [O II] is intrinsically weak in the optical spectra, particularly in quasars, and the degree to which the AGN itself contributes to the observed [O II] emission remains uncertain \citep[e.g.][]{oii_ho2005, maddox2018,zhuang2019}.

Recently, flux ratios involving [O II] have been frequently employed to assess the relative importance of star formation in AGN host galaxies. In particular, the [O II]/[Ne V] ratio has gained traction as a diagnostic because [Ne V] $\lambda$3426 is exclusively excited by AGN activity and not produced by star formation, making it a useful indicator of the relative contributions of star formation and AGN emission to the observed [O II] flux \citep[e.g.][]{maddox2018,Vergani2018,zfchen2022}. Meanwhile, the [O II]/[O III] ratio has also been widely used for a similar purpose \citep[e.g.][]{oii_ho2005,Silverman2009,zhuang2019}.

A key physical distinction between [O II] and the high-ionization lines [Ne V] and [O III] is that the low critical density of [O II] \citep[critical density $\sim10^3-10^4\,\mathrm{cm}^{-3}$,][]{Osterbrock2006,Zheng1988} restricts its production to low-density gas at large scales, resulting in a relatively narrow line profile. By contrast, [Ne V] and [O III] can arise in the higher-density, more compact regions of the NLR, which may also be significant outflowing, leading to characteristically broader and often blueshifted profiles \citep[e.g.,][]{oiii_broad_peng2014, oiii_broad_singha2022}.


Therefore, by comparing the line profiles of [O II] with those of these high-ionization lines in high-S/N spectra, one can effectively isolate the [Ne V] or [O III] emission associated with the low-density, large-scale NLR that produces [O II], separating it from emission produced in denser and likely outflowing NLR gas that contributes little to [O II]. This thus could provide a novel and kinematically-consistent approach to determine physically meaningful [O II]/[Ne V] and [O II]/[O III] flux ratios, which can be directly compared with CLOUDY calculations for low-density gas, yielding more accurate constraints on the AGN contribution to [O II] emission.

In this work, we construct composite spectra of large quasar samples to obtain high-S/N emission-line profiles for [O II] and [Ne V], and systematically investigate the AGN contribution to [O II] emission. A parallel comparison between [O II] and [O III] is also performed as a complementary check (see \S\ref{sec:discussionA}). The paper is organized as follows. In \S\ref{sec:sample},we describe the sample selection, the construction of the composite spectra, and the spectral fitting procedures. In \S\ref{sec:3}, we present our line-decomposition methodology and demonstrate that the AGN contribution to the \OII\ emission is rather low in quasars. In \S\ref{sec:sf}, we examine how the \OII-based star formation rate evolves with AGN luminosity in quasars, and present a discussion of the results in \S\ref{sec:discussion}.

\begin{figure*}[htbp]
    \centering
    \includegraphics[width=0.9\textwidth]{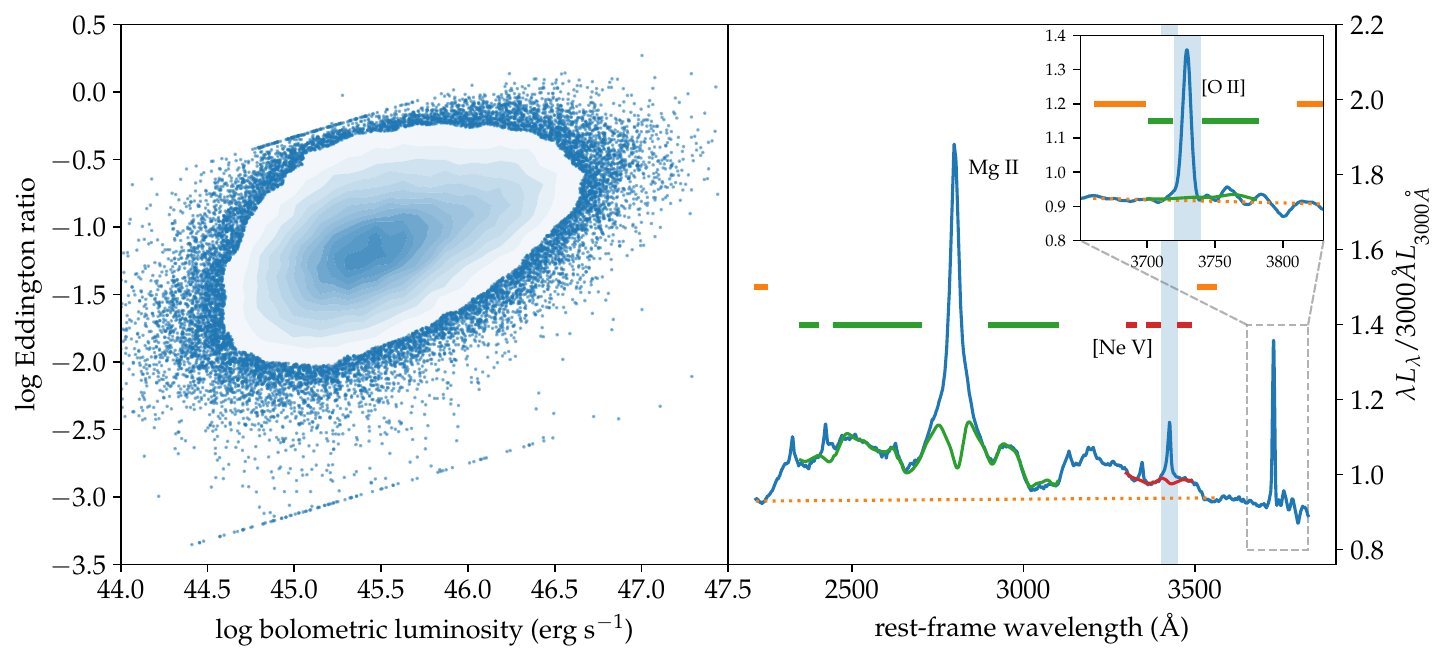}
    \caption{\textbf{Sample and spectral fitting methodology.}
        \textbf{Left panel:} Distribution of the full sample in the $L_{\rm bol}$--$\lambda_{\rm Edd}$ parameter space. 
        The apparent linear features formed by data points in the upper-left and lower-right corners arise from the truncation applied to the Mg\textsc{ii} FWHM values in \citet{ws22}, and the affected sources represent only a negligible portion of the sample.
        \textbf{Right panel:} The composite spectrum of the full sample illustrating the continuum and Fe II subtraction procedure. 
        In the main panel, the Fe II fitting includes three free parameters: the velocity dispersion and two independent amplitudes for the wavelength ranges 2200–3100 \AA\ and 3250–3500 \AA\, respectively, with the interval 3100–3250 \AA\ excluded from the fit \citep{zfchen2022}. 
        The thick orange horizontal line indicates the fitting window for the power-law continuum. Green and red lines mark the wavelength ranges for two separate iron templates with independent amplitudes. The fitted continuum and iron components are shown in corresponding colors. The [O II] $\lambda\lambda$3726,3729 doublet region is indicated by a gray dashed box in the main panel, with a zoomed-in view shown in the upper-right inset where a single iron template (green line) is used.
    }
    \label{fig:sample_and_method}
\end{figure*}

\section{Sample, spectra stacking and fitting}\label{sec:sample}

We begin with the Sloan Digital Sky Survey (SDSS) Data Release 16 quasar catalogue \citep{lyke20}, which contains 750,414 spectroscopically confirmed quasars and represents the largest such compilation from SDSS to date. To ensure rest-frame spectral coverage over 2220–3830 \AA\ (see Fig.~\ref{fig:sample_and_method}), we restrict the sample to the redshift range $0.80 \le z \le 1.35$. We further require available bolometric luminosity and Eddington ratio measurements from \citet{ws22}, yielding a final sample of 171,233 quasars, whose distribution in the $L_{\rm bol}$–$\lambda_{\rm Edd}$ plane is shown in the left panel of Fig.~\ref{fig:sample_and_method}.

Because weak emission lines such as \OII\ and \NeV\ are poorly constrained in most individual spectra, we therefore rely on composite spectra. To construct the composites, individual spectra are first normalized to unit flux density within a 20 \AA\ continuum window centered at 3000 \AA, after which the composite spectrum is obtained by averaging the normalized spectra. 
The stacked spectrum of the full sample over the rest-frame 2220–3830 \AA\ range, which is adopted for the subsequent spectral fitting,
is shown in Fig.~\ref{fig:sample_and_method}.

Before measuring emission-line properties, we fit and subtract the continuum and Fe II emission. Following \citet{zfchen2022}, the continuum is modeled as a power law, while Fe II emission is represented using empirical templates from \citet{fe2_iwz1} and \citet{fe2_tsuzuki06}. 
The fitting procedure is illustrated in Fig.~\ref{fig:sample_and_method} using the composite spectrum of the full sample.

\section{Line decomposition}\label{sec:3}




\begin{figure}[htbp]
    \centering
    \includegraphics[width=0.48\textwidth]{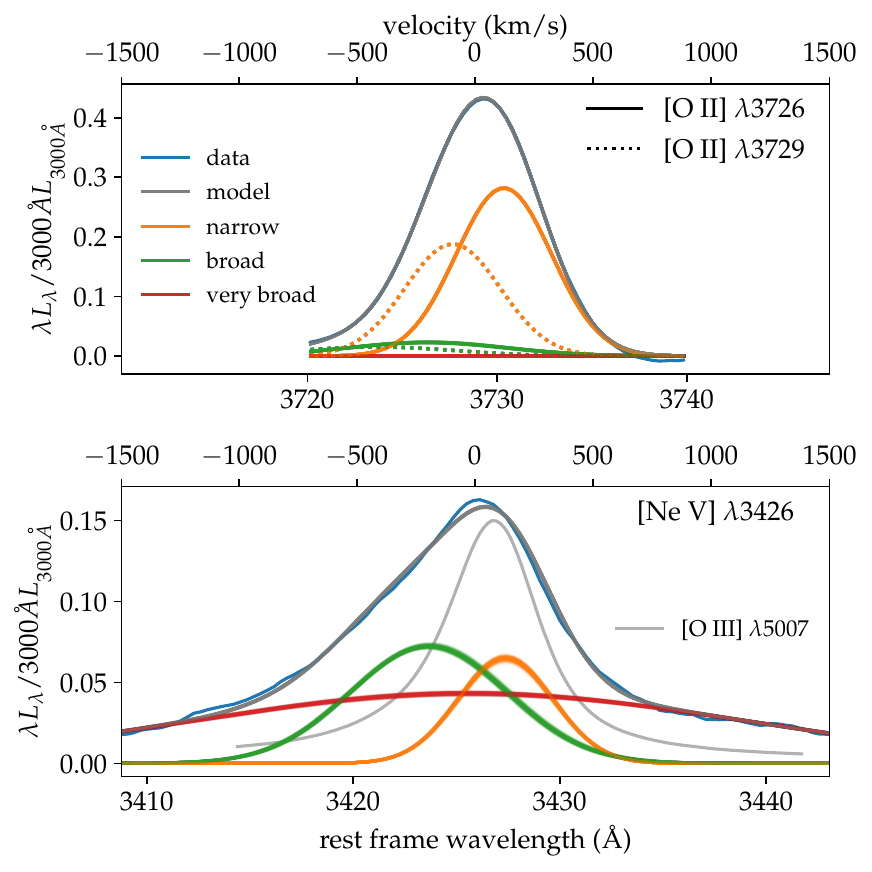}
    \caption{\textbf{Kinematically consistent decomposition of [O II] and [Ne V] emission lines for the full sample.}
    Both panels show dual x-axes in wavelength and velocity, with the velocity axes aligned for direct comparison.
    The observed profile is shown in blue, and each line is decomposed into narrow (orange), ``broad (green)", and ``very broad" (red) Gaussian components, whose widths and velocity shifts are tied between \OII\ and \NeV. In the \OII\ doublet (upper panel), the $\lambda3726$ and $\lambda3729$ lines are indicated by dotted and solid curves, respectively. For each component and for the total model (gray), 100 randomly drawn MCMC posterior samples are overplotted as semi-transparent curves to illustrate the fitting uncertainties.
    In the lower panel, a scaled [O~III] emission-line profile (light gray) is overplotted as a reference (see \S\ref{sec:discussionA}).
    }
    \label{fig:fitting_full_sample}
\end{figure}

We perform Markov Chain Monte Carlo (MCMC) analysis with 10,000 iterations to simultaneously fit the \OII\ and \NeV\ emission lines derived from the composite spectra.
As shown in Fig.~\ref{fig:fitting_full_sample}, the composite \OII\ line profile of the full sample is primarily described by a single narrow Gaussian component for each line of the \OII\ $\lambda\lambda3726,3729$ doublet, with the doublet separation fixed at the theoretical value of 224 km/s and the flux ratio F($\lambda3729$)/F($\lambda3726$) fixed at 1.5, corresponding to the low-density limit \citep{pradhan2006}.

In contrast, \NeV\ $\lambda3426$ line is significantly broader and more blueshifted. Fitting the \NeV\ line requires, in addition to the narrow component (with its line width and velocity shift tied to those of \OII), two extra broader and blueshifted Gaussian components, referred to as the “broad” and “very broad” components. For methodological consistency, these ``broad" and ``very broad" components are also included in the \OII\ fitting; taken together, they contribute only a minor fraction (12\%) of the total \OII\ emission. From the MCMC posterior distributions, we derive the line flux/luminosity of each component.

The total \OII/\NeV\ flux ratio measured from the full composite spectrum is 1.331$\pm$0.003. Such a low ratio would, if interpreted at face value, leave little room for a significant contribution of star formation to the observed \OII\ emission in quasars \citep{oii_ho2005}. However, once the line profiles are decomposed into kinematically consistent components, a markedly different picture emerges. The \OII/\NeV\ flux ratio associated with the narrow component is substantially higher (7.86$\pm$0.13), whereas that for the combined “broad” and “very broad” components is much lower (0.184$\pm$0.003).

The blueshifted ``broad" and ``very broad" components are naturally interpreted as arising from high-density, highly ionized outflowing gas, in which little \OII\ emission is expected to be produced. In contrast, the \OII\ emission is almost entirely dominated by the narrow component, rendering only the narrow-component \OII/\NeV\ flux ratio physically meaningful for diagnosing star formation. The resulting ratio is significantly higher than the value expected for low-density, AGN radiation-pressure–dominated narrow-line region clouds (\OII/\NeV\ $\simeq 0.5$; \citealt{zhuang2019,zfchen2022}), indicating that the observed \OII\ emission in quasars is dominated by star formation. 

Based on the observed flux of the narrow \NeV\ component and adopting an AGN-origin narrow \OII/\NeV\ flux ratio of 0.5 \citep{zhuang2019, zfchen2022}, we estimate that the AGN contributes only 6.4\% of the observed narrow \OII\ component in the full sample. This fraction would be higher (17.4\%) if the weak ``broad" and ``very broad" components of \OII\ emission, which are most plausibly associated with AGN activity, were included in the total \OII\ emission. Therefore, the observed \OII\ emission can be used as a probe of star formation in quasars with only a minor correction for AGN contamination.

{We finally note that the method presented here can also be applied to individual sources to assess the AGN contribution to \OII\ emission, provided that high-quality measurements of the \OII\ and \NeV\ line profiles are available. When such an assessment is not feasible, directly assuming that the \OII\ emission is dominated by star formation should be treated with caution. The relative contributions of star formation and AGN may vary from object to object, and in some cases the \OII\ emission could be significantly contaminated, or even dominated, by the AGN narrow-line region, although our stacking results suggest that such cases should represent only a minority of the quasar population.}


\section{The star formation rates in quasars}\label{sec:sf}

\begin{figure}[htbp]
    \centering
    \includegraphics[width=0.48\textwidth]{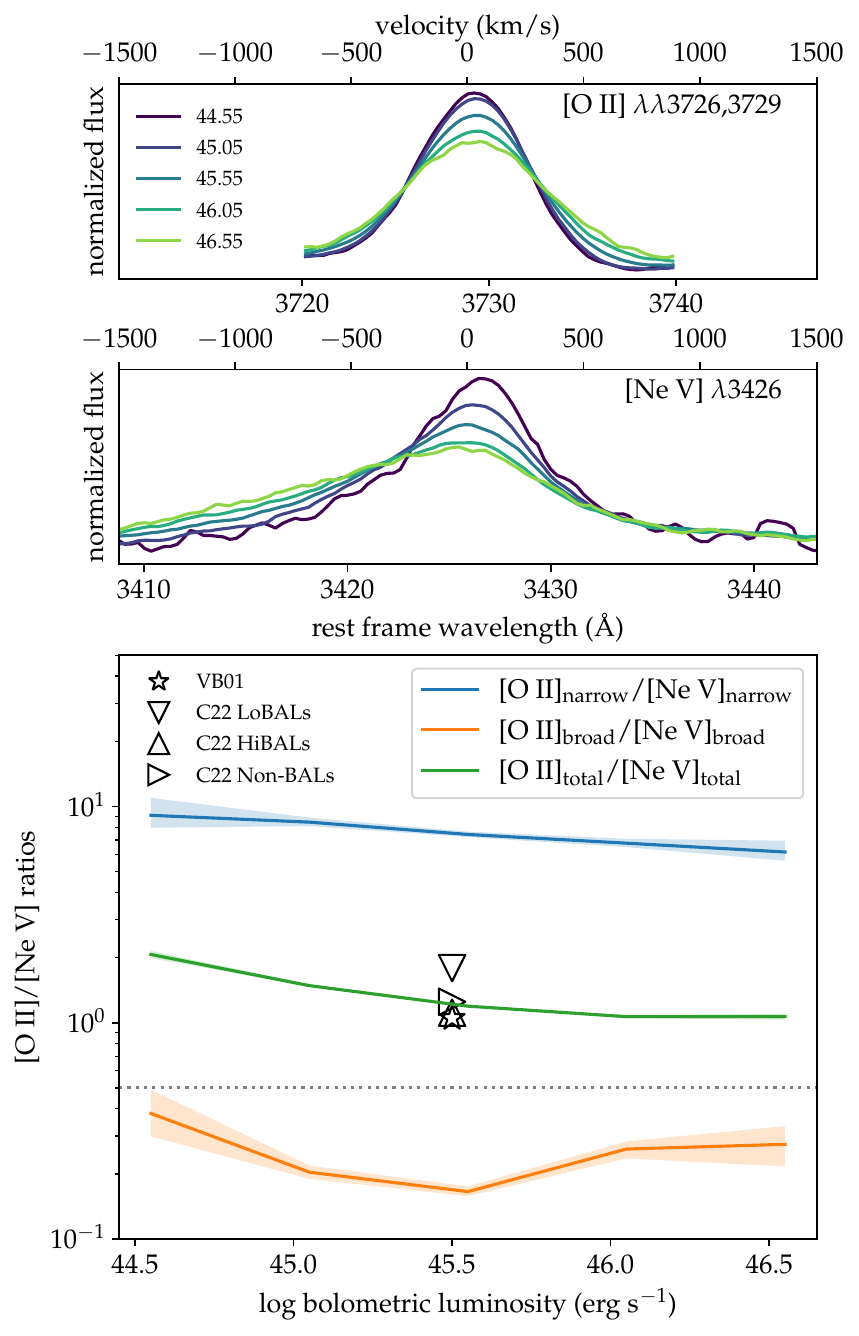}
    \caption{\textbf{\OII\ and \NeV\ line profiles and flux ratios as a function of bolometric luminosity.} 
        \textbf{Upper and middle panels:} Observed \OII\ and \NeV\ line profiles in five bolometric-luminosity bins, normalized to the same integrated flux to facilitate comparison of line widths and shapes.
        \textbf{Lower panel:} [O II]/[Ne V] flux ratios for narrow component (blue), broad components (orange), and total flux (green). Previous total-flux measurements from \cite{vdb2001} (star) and \cite{zfchen2022} (triangles) are shown for comparison. The horizontal dotted line marks the flux ratio of 0.5 predicted by CLOUDY for AGN photoionization.
    }
    \label{fig:profiles}
\end{figure}

To investigate how the \OII-traced star formation rate evolves with quasar luminosity, we bin the full sample into five intervals of bolometric luminosity, spanning $44.3 \le \log L_{\rm bol} < 46.8$ (erg/s) in steps of 0.5 dex. This binning provides sufficient signal-to-noise in each composite spectrum while preserving sensitivity to potential luminosity-dependent trends. The derived \OII\ and \NeV\ line profiles and the line ratio as a function of luminosity are shown in Fig.~\ref{fig:profiles}. The observed narrow \OII/\NeV\ flux ratios, much higher than the corresponding total flux ratios, further indicate that the \OII\ emission in each luminosity bin is dominated by star formation.


After correcting for the minor AGN contribution, we convert the star-formation–related \OII\ luminosities into star formation rates using the calibration ${\rm SFR}(M_{\odot}~{\rm yr}^{-1})=(1.4 \pm 0.4) \times 10^{-41} L_{\rm [OII]}({\rm erg~s}^{-1})$ \citep{kennicutt1998}.
Note that the [O II]-based SFR calibration depends on oxygen abundance \citep[e.g.][]{kewley2004}. However, the stellar mass--metallicity relation flattens at the high-mass end ($M_* \gtrsim 10^{10.5} M_\odot$) \citep[e.g.][]{tremonti2004}, where most of our quasar hosts reside. The characteristic abundance at this high-mass end, 12 + log(O/H) $\sim$ 9.1, when substituted into the \citet{kewley2004} calibration yields a coefficient consistent with that of \citet{kennicutt1998}.

\begin{figure}[htbp]
    \centering
    \includegraphics[width=0.48\textwidth]{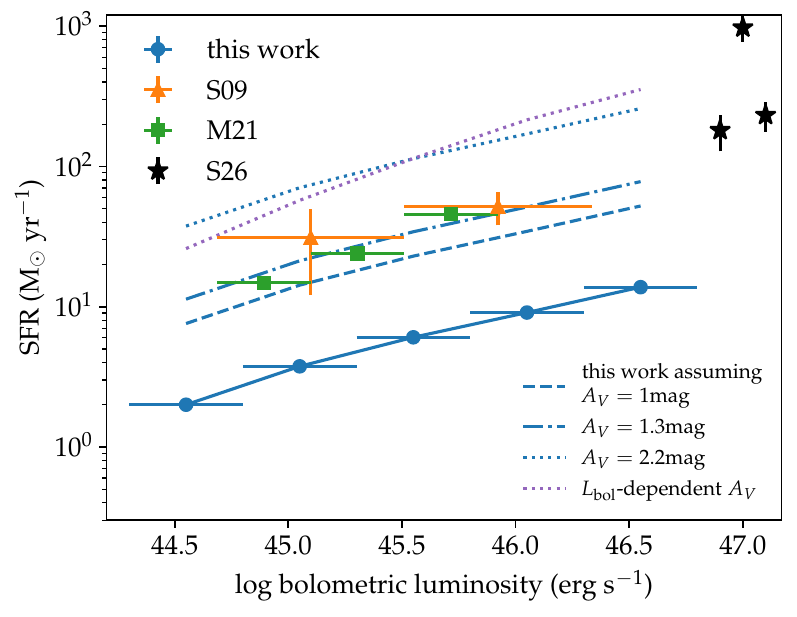}
    \caption{\textbf{Star formation rates as a function of quasar bolometric luminosity.}
        Star formation rates derived from host [O II] luminosity are shown without dust correction (blue circles), and with constant dust corrections of $A_V = 1.0$ (dashed line), $1.3$ (dash-dotted line), and $2.2$ (blue dotted line) mag, as well as with a bolometric-luminosity-dependent dust correction based on the $M_{\rm BH}$–$M_\star$ relation (purple dotted line). Horizontal error bars indicate the bin width. For comparison, independent SFR estimates for quasar hosts from far-infrared measurements \citep[$z \sim 0.5$–$1$, orange triangles;][]{serjeant2009}, radio continuum measurements \citep[$z \sim 0.8$–$1.2$, green squares;][]{macfarlane2021}, and {CO ($J=5$–$4$) measurements for three luminous quasars at $z = 2$ \citep[black stars;][]{silverman2026} are also shown.}
    }
\label{fig:ratios}
\end{figure}

Fig.~\ref{fig:ratios} compares our SFR measurements with independent estimates from the literature at comparable redshifts,
enabling meaningful comparison.
Our [O II]-based SFRs are systematically lower by a factor of $\sim$4 compared to the FIR and radio estimates. This discrepancy could arise from several factors: (1) dust extinction suppressing the observed [O II] flux, (2) AGN contamination in the FIR or radio measurements leading to overestimated SFRs \citep[e.g.,][]{symeonidis2016}, or (3) the limited aperture of the SDSS fiber missing extended star formation captured by FIR and radio observations.

{
Following \citet{oii_ho2005}, we first adopt an extinction correction of $A_V = 1.0$ mag, a typical value for star-forming galaxies. As an independent estimate, we use the typical SMBH mass of our quasar sample to infer the host galaxy stellar mass based on the $M_{\rm BH}$–$M_\star$ relation \citep{greene2020}. For the typical SMBH mass of our sample ($M_{\rm BH} \sim 10^{8.5}M_\odot$), this relation implies a stellar mass of $M_\star \sim 10^{11.1}M_\odot$. The expected dust attenuation can then be estimated using the $A_V$–$M_\star$ relation of \citet{garn2010}, which predicts $A_{\mathrm{H}\alpha} \sim 1.8$ mag at $M_\star \sim 10^{11.1}M_\odot$. Using the attenuation curve of \citet{calzetti2000}, this corresponds to $A_V \sim 2.2$ mag.

Motivated by this relation, we also estimate the dust attenuation for each luminosity bin in Fig.~\ref{fig:ratios} by adopting the median SMBH mass in each bin, converting it to stellar mass, and then deriving the corresponding extinction from the $A_V$–$M_\star$ relation. 

Using the different $A_V$ assumptions described above, we show the resulting extinction-corrected SFRs for our sample in Fig.~\ref{fig:ratios}.

We find that while the derived extinction-corrected SFR–$L_{\rm bol}$ relation depends on the assumed dust extinction, adopting a constant value of $A_V \sim 1.3$ mag yields results that are in good agreement with previous studies in the literature (Fig.~\ref{fig:ratios}). This value is slightly higher than the commonly adopted $A_V = 1.0$ mag for star-forming galaxies, likely reflecting the higher stellar masses of quasar host galaxies. However, it remains lower than the value predicted from the $A_V$–$M_\star$ relation, which may indicate that quasar activity reduces the dust extinction in their host galaxies. In the following discussion, unless otherwise stated, we interpret our results assuming a constant extinction correction of $A_V \sim 1.3$ mag.

}


We note that an accurate and independent determination of dust extinction affecting the \OII\ emission is however not feasible for our sample. This is not only because the traditional Balmer-decrement method requires spectral coverage of both H$\alpha$ and H$\beta$, but also because the dust extinction toward the narrow-line region and the star-forming regions in quasar host galaxies may differ substantially. 


\section{Discussion}\label{sec:discussion}

\subsection{AGN Contribution to the \OII\ Emission}\label{sec:discussionA}



We repeat the analysis described in \S\ref{sec:3}, but using the \OII/\OIII\ diagnostic instead. For this purpose, we select quasars within the redshift range of $0.1 \le z \le 0.75$, where both emission lines are covered. Adopting the expected AGN-origin \OII/\OIII\ flux ratio of 0.108 \citep{zhuang2019}, we find that the AGN contributes $\sim$ 46\% of the observed \OII\ emission (34\% of the narrow component) in quasars. This fraction is considerably higher than the 17.4\% (6.4\%) derived from the \OII/\NeV\ diagnostic. This discrepancy likely arises because \NeV\ has both a higher ionization potential (97 eV versus 35 eV) and a higher critical density (1.6$\times$10$^7$ vs 7.0$\times$10$^5$ cm$^{-2}$, \citealt{DeRobertis1984}) than \OIII. As a result, \NeV\ emission is more strongly dominated by gas in the inner, high-density, and outflowing regions of the AGN, for which the kinematically consistent line-decomposition approach is more effective (see Fig. \ref{fig:fitting_full_sample} for a comparison with \OIII). Moreover, \OIII\ emission may receive (weak) contribution from star-forming regions in the host galaxy. Another advantage of the \OII/\NeV\ diagnostic  is that they have more closely spaced wavelengths, making their flux ratio largely insensitive to dust reddening in the NLR. Therefore, the results based on the \OII/\NeV\ diagnostic are expected to be more reliable, while the AGN contribution inferred from the \OII/\OIII\ diagnostic should be regarded as an upper limit.



Notably, \citet{zhuang2019} reported a median AGN contribution of 13.8\% to the \OII\ emission for a sample of SDSS narrow-line AGN, based on the \OII/\OIII\ diagnostic, which is broadly consistent with our conclusion that the observed \OII\ emission is dominated by star formation. However, the numerical value of this fraction cannot be directly compared with our measurements for several reasons.
First, their sample is drawn from the SDSS galaxy catalog and is restricted to low-redshift AGN (z $<$ 0.4), whereas our sample consists of quasars at significantly higher redshifts. Second, unlike in our analysis, no emission-line profile decomposition was performed in their work, which is likely to overestimate the AGN contribution. Third, their emission-line fluxes were corrected for dust extinction using the Balmer decrement, which was not applied to our sample and would, if applied, likely result in a lower inferred AGN contribution fraction, given that the \OII\ emission is more strongly affected by dust extinction than \OIII. Finally, their AGN selection requires significant detections of the \OII\ line in individual sources, which inevitably biases the sample toward \OII-strong sources and consequently toward lower inferred AGN contribution fractions. We finally stress that when investigating the \OII\ emission in specific AGN populations or in individual sources, a tailored analysis is essential in order to obtain a reliable estimate of the AGN contribution to the observed \OII\ emission.


\subsection{Evolution of Star Formation in Quasars}\label{sec:evolution}

A large body of observational work has investigated the relation between the star formation rate in AGN host galaxies and the AGN intrinsic luminosity (see \citealt{Harrison2017,Alexander2012,Alexander2025} for reviews). Many of these studies, particularly those based on X-ray–selected AGN samples, find that the observed relation is relatively flat, at least at low to moderate AGN luminosities \citep[e.g.,][]{Silverman2009, Stanley2015}.
This apparent lack of strong correlation is now widely interpreted as a consequence of the strong short-timescale variability of AGN X-ray emission, which can substantially dilute the underlying connection when compared with star-formation tracers that probe much longer timescales \citep[e.g.,][]{Hickox2014}.

In this work, we find that the [O\,\textsc{ii}]-based SFR in quasar host galaxies strongly correlates with quasar luminosity, following a relation of the form 
$\mathrm{SFR} \propto L_{\mathrm{bol}}^{\alpha}$, with $\alpha = 0.44\pm0.02$.
The derived value of $\alpha$ is in good agreement with previous results obtained from large quasar samples using far-infrared or radio-based SFR measurements \citep[e.g.,][]{serjeant2009, dong2016, macfarlane2021}. 
The relatively steep slopes consistently found in these optical-selected quasar samples, compared with the much flatter relations reported for X-ray–selected AGN samples, may in part reflect the fact that quasar optical/UV emission exhibits substantially weaker short-timescale variability than X-ray emission in low- to moderate-luminosity AGN. 




Thanks to our large sample size and the high signal-to-noise composite spectra, which enable precise [O II]-based SFR measurements, we are able, for the first time, to disentangle the respective roles of the two fundamental physical parameters underlying the luminosity, namely the SMBH mass and the Eddington ratio, in driving the observed SFR--luminosity correlation in quasars.

To this end, we perform a residual analysis to disentangle the effect of each parameter. Specifically, we carry out linear regressions to model: (1) $\log L_{\rm bol}$ as a function of $\log M_{\rm BH}$; (2) $\log L_{\rm bol}$ as a function of $\log \lambda_{\rm Edd}$.
For each regression, we compute the residuals, which represent the component of the dependent variable that is independent of the corresponding independent variable. We then bin the sample according to these residuals into five bins with a width of 0.5 dex, spanning the range from -1.25 to 1.25. For each bin, we construct composite spectra and apply the kinematically consistent emission-line fitting procedure described in \S\ref{sec:3}. 

We find 
$\mathrm{SFR} \propto M_{\mathrm{BH}}^{0.50\pm0.01}$ 
at fixed $\lambda_{\rm Edd}$ and 
$\mathrm{SFR} \propto \lambda_{\rm Edd}^{0.46\pm0.02}$ at fixed $M_{\rm BH}$, respectively. 
The positive correlation between SFR and $\lambda_{\rm Edd}$ 
can be naturally interpreted within the framework in which both star formation and SMBH accretion are regulated by a common cold-gas reservoir \citep[e.g.][]{mullaney2012,shangguan2020}. 
In this picture, an enhanced gas supply simultaneously fuels stellar growth and black hole accretion. 
Alternatively, AGN-driven \emph{positive feedback} \citep[e.g.][]{hopkins2006,zhuang2021,xie2021} may also contribute to this correlation, 
whereas scenarios dominated by \emph{negative feedback} are disfavored by the observed trend.  Meanwhile, the relation 
$\mathrm{SFR} \propto M_{\mathrm{BH}}^{0.50\pm0.01}$ at fixed $\lambda_{\rm Edd}$ suggests that, in more massive host systems, sustaining the same normalized accretion rate onto the SMBH requires a higher overall fuel supply, as traced by an elevated star formation rate.


Interestingly, after performing a regression to model $\log \lambda_{\rm Edd}$ as a function of $\log L_{\rm bol}$, obtaining the corresponding residuals, and rebinning the sample, we find
$\mathrm{SFR} \propto \lambda_{\rm Edd}^{-0.11\pm0.01}$ (at fixed $L_{\rm bol}$, and subsequently,  $\mathrm{SFR} \propto M_{\rm BH}^{0.11\pm0.01}$), 
indicating a weak dependence of SFR on $\lambda_{\rm Edd}$ (or $M_{\rm BH}$) once the bolometric luminosity is controlled. 


The above analysis indicates that, for quasars, the bolometric luminosity is the primary parameter most tightly linked to the observed mean star formation activity in their host galaxies, supporting the scheme that both quantities may be jointly regulated by the availability of cold gas.

Our findings complement the long-term coevolution framework established by \citet{mullaney2012}, who found a constant $\langle \dot{M}_{\rm BH} \rangle / \langle {\rm SFR} \rangle$ ratio {$\sim(0.6\text{--}0.8)\times10^{-3}$} by averaging over star-forming galaxy populations at $z \sim 1$--2. While their constant ratio reflects cosmological coevolution over duty cycles, including predominantly low-luminosity or quiescent systems, our observed SFR--$L_{\rm bol}$ correlation probes the physical connection during high-power episodes ($\langle \dot{M}_{\rm BH}\rangle \sim L_{\rm bol} = 10^{44-47}~{\rm erg~s^{-1}}$, roughly 1--2 orders of magnitude more luminous than their average). {In our quasar sample, the $\dot{M}_{\rm BH}/{\rm SFR}$ ratio increases from $\sim8\times10^{-3}$ to $\sim0.12$ across the five luminosity bins, 1--2 orders of magnitude above the $\sim(0.6\text{--}0.8)\times10^{-3}$ reported by \citet{mullaney2012}, reflecting the elevated instantaneous accretion rates during quasar phases. This is further supported by the three luminous quasars at $z=2$ ($L_{\rm bol} \sim 10^{47}\mathrm{erg/s}$) studied by \citet{silverman2026}, whose $\dot{M}_{\rm BH}/{\rm SFR}$ ratios of $\sim0.02$--$0.11$ are broadly consistent with an extrapolation of our results to higher luminosities, although there is substantial source-to-source variation.} Meanwhile, given the substantial variability expected during such high-power phases, it is therefore not surprising that we observe an SFR--$L_{\rm bol}$ relation ($\alpha$ = 0.44) significantly flatter than 1.0.

Importantly, we further find that the SFR--$L_{\rm bol}$ relation in quasars exhibits rather weak dependence on SMBH mass. When considered together with the results of \citet{mullaney2012}, who reported that the ratio $\langle \dot{M}_{\rm BH} \rangle / \langle {\rm SFR} \rangle$ shows little dependence on the stellar mass of the host galaxies\footnote{This result may hold primarily for massive galaxies; subsequent studies extending to lower-mass systems have found a clear stellar-mass dependence\citep[e.g.][]{Yang2018, Torbaniuk2024}.},
this suggests that not only the time-averaged growth of SMBHs over cosmic duty cycles, but also the instantaneous accretion power attained during quasar phases, is closely tied to the star formation activity of the massive host galaxy, and hence to the availability of cold gas, with only a weak dependence on the mass of the system. 

{A final note is that the discussion in this section is based on \OII-derived SFR estimated assuming a constant dust extinction correction. Assuming a stellar mass dependent (thus SMBH mass dependent) extinction correction would yield steeper SFR dependence on luminosity (see the purple dotted line in Fig. \ref{fig:ratios}) or mass. For example, the inferred relations would become SFR $\propto L_{\rm bol}^{0.58}$ (instead of $L_{\rm bol}^{0.44}$), SFR $\propto M_{\mathrm{BH}}^{0.71}$ (instead of $M_{\mathrm{BH}}^{0.50}$ at fixed $\lambda_{\rm Edd}$), and SFR $\propto M_{\mathrm{BH}}^{0.44}$ (instead of $M_{\mathrm{BH}}^{0.11}$ at fixed $L_{\rm bol}$).
Under this assumption, the results would suggest that, in addition to luminosity, the SMBH mass is also positively correlated with the SFR in quasar host galaxies at fixed luminosity. This would modify our previous conclusion that the SFR exhibits only a very weak dependence on SMBH mass at fixed luminosity. However, this possibility should be treated with caution. The SMBH mass measurements themselves carry substantial uncertainties, and additional uncertainties arise from the conversion from SMBH mass to stellar mass and from stellar mass to dust extinction. Furthermore, quasar activity may also affect the dust content in their host galaxies. Therefore, the robustness of this alternative interpretation remains uncertain. Further observational constraints on dust attenuation in quasar host galaxies will be important to better assess this possibility.

}

\section*{Acknowledgements}
{We thank the anonymous referee for constructive comments and suggestions that have improved the quality of this work.}
This work was supported by National Key R\&D Program of China (No. 2023YFA1607903 and 2023YFA1607904),
the Chizhou University High-level Talent Research Start-up Fund (CZ2025YJRC104), the National Natural Science Foundation of China (grant nos. 12533006,  12192221), and Guizhou Provincial Major Scientific and Technological Program XKBF (2025)010 and XKBF (2025)011. LCH acknowledges support from the National Natural Science Foundation of China (12233001) and the China Manned Space Program (CMS-CSST-2025-A09).

\bibliography{main}
\bibliographystyle{aasjournal}



\end{document}